\documentclass[a4paper,twoside]{article}

\usepackage{epsfig}
\usepackage{subfigure}
\usepackage{calc}
\usepackage{amssymb}
\usepackage{amstext} 
\usepackage{amsthm}
\usepackage[table,xcdraw]{xcolor}
\usepackage[fleqn]{amsmath} 
\usepackage{pslatex}
\usepackage{enumitem}
\usepackage{textcomp}
\usepackage{soul}
\usepackage{natbib}
\usepackage{array,multirow,graphicx}

\usepackage{SCITEPRESS}     % Please add other packages that you may need BEFORE the SCITEPRESS.sty package.

\begin{document}

\title{Cultural Influences on Requirements Engineering Process\\ in the Context of  Saudi Arabia} 
%Cultural Influences on Requirements Engineering Process in the Context of  Saudi Arabia

 \author{\authorname{Tawfeeq Alsanoosy, Maria Spichkova, James Harland}
 \affiliation{School of Science, RMIT University, Melbourne, Australia} 
 \email{\{tawfeeq.alsanoosy, maria.spichkova, james.harland\}@rmit.edu.au
 }}


\abstract{ Software development requires intensive communication between the requirements engineers and software stakeholders, particularly during the Requirements Engineering (RE) phase.   
Therefore, the individuals' culture might influence both the RE process and the result. 
Our aims are to investigate the extend of cultural influences on the RE process, and to analyze how the RE process can be adapted to take into account cultural aspects. 
The model we present is based on Hofstede's cultural theory. 
The model was applied on a pilot case study in the context of the conservative Saudi Arabian culture.   
The results reveal 6 RE aspects and 10 cultural factors that have a large impact on the RE practice. 
 }

\onecolumn 
\maketitle 
\normalsize \vfill

\section{\uppercase{Introduction}}
\label{sec:introduction}

\noindent 
Requirements Engineering (RE) is one of the key phases of the software development life cycle. 
RE refers to the process of identifying and documenting the stakeholder's requirements. 
It was initially introduced for use within Western culture before the expected benefits began attracting the RE practitioners in various other cultures,  
to adopt the RE process.  
The RE process has a critical impact on software quality.  
It requires an intensive communication between the requirements engineers and software stakeholders, in order to elicit, validate, and refine the requirements. 
This requires comprehensive communication skills, and an understanding of the individuals' behavior and culture \citep{thanasankit2002requirements,hanisch2007impediments}.

Culture has a great influence on how individuals and companies operate and how they adopt techniques, methods, and practices to achieve their goals.   
Culture shapes the way in which people think, communicate, understand, and select what is important \citep{hofstede2010cultures}. 
Cultures have distinctive beliefs, customs, and approaches to communication.
This diversity is influenced by the behavior practice within these cultures.
\cite{hanisch2001understanding} argued that the social and cultural aspects of RE affect the success of software development and it cannot be ignored.
The objectives of our research are to identify the cultural influences on the RE processes and practices, and to analyze how the RE process can be adapted to take into account cultural aspects.

But what do we usually understand under the term \textit{culture}?
There are several definitions for it, specified from the social psychology view. 
Hofstede, whose work has been cited 70,392 times\footnote{According to the Google Scholar, retrieved 3/11/2017 
}, defines \textit{culture} as\\\textit{{\textquotedblleft}collective programming of the mind which distinguishes members of one human group from another{\textquotedblright}},\\ cf. \citep{hofstede2010cultures}. 
We take this approach as a background for our research not only because  
Hofstede conducted one of the most inclusive studies on the influence of cultural values on the workplace, but also because this is the most SE-related social psychology study. 
Hofstede conducted the research with employees working for IBM, covering more than 50 countries.
This theory is widely accepted and adopted in software engineering cultural studies \citep{Borchers,lim2015investigating,ayed2017agile}.

\textbf{Contributions:} The RE process is originated from Western culture and hence most of the standard RE practices are driven by the United States. 
Successfully applying such process/practice within different cultural context would, therefore considering the cultural differences.
This study investigates the impact of the Saudi national culture on the practice of RE process. 
We collected data, through interviewing 6 software practitioners and 2 software engineering researchers, about the RE practices that were less or more successfully adapted within the Saudi Arabian culture.
We then identified a model designed to illustrate the interplay between the core of the RE process and the Hofstede cultural dimensions, within the Saudi culture. %  
The results demonstrated the feasibility of our hypothesis as well as allowed us to identify a missing hypothesis, which we added to the refined model.
The current results of this study might be useful for improving the RE process within the Saudi culture and any culture that has similar national cultural profiles.

%=======================================================

 \textbf{Outline:} This paper is structured as follows. 
Section \ref{sec:Background} introduces the background of our research, 
focusing on RE processes and Hofstede's cultural theory.
Our conceptual Model and hypotheses are presented in Section \ref{sec:Conceptual_Model}.
Section \ref{sec:study} describes the research methodology.
The results of the case study are presented in Section \ref{sec:Result}. 
Section \ref{sec:model} introduces the proposed framework. 
In Section \ref{sec:RELATED}, the related work is discussed.
Section \ref{sec:conclusion} summarizes the paper. 

%=======================================================================
\section{\uppercase{Background}}
\label{sec:Background}

%------------------------------------------------
\subsection{Requirements Engineering Process} 
\label{sec:Bckground_RE}

\cite{zave1997classification} defines RE as 
\textit{{\textquotedblleft}the branch of software engineering that is concerned with the real-world goals for, functions of, and constraints on software systems. It is also concerned with the relationship of these factors to precise specifications of software behavior, and to their evolution over time and across software families{\textquotedblright}}.
As highlighted by \citep{nuseibeh2000requirements}, the above definition
features  three important aspects of RE:
(1) RE is concerned with real-world goals,
(2) its aim is to provide precise specifications of the requirements, and
(3) the definition emphasizes the reality of the rapid change of user's needs.
Thus, requirements specification serves as a basis for the development of the system, its design and architecture, cf. \cite{spichkova2011architecture}.
In this paper, the term \textit{requirements engineer} is used to refer to the practitioner who works on specifying, analyzing, and documenting users' requirements.     

The RE process is usually described as the following five core (sub-)processes/activities, cf. also \cite{SommervilleIan2004Se,abran2004guide,pandey2010effective}: 
\begin{itemize}
    \item
    \textbf{Requirements elicitation:} The process  of gathering, identifying, extracting, and exposing the stakeholders' needs;  
    \item 
    \textbf{Requirements analysis:} The process of refining stakeholders' requirements and constraints by defining the procedure, input/ output data and object of the required system; 
    \item
    \textbf{Requirements specification:} The process of creating a written description of stakeholders’ requirements, needs and constraints in a complete, accurate, understandable and consistent format;
    \item
    \textbf{Requirements validation:} The activity to ensure that the requirement specification is consistent, clear and capture users' needs and constraints;
    \item
    \textbf{Requirements management:} The process of tracing changes to the requirements over the time of system development and, in some cases, application as well.
\end{itemize}

%------------------------------------------------
\subsection{Hofstede's Cultural Theory}
\label{sec:Background_Hofstede}

\cite{hofstede2010cultures} proposed to focus on the following six dimensions of a national culture:
\begin{itemize}
\item 
\textbf{Power Distance Index (PDI):} The degree to which the lesser authority individual accepts and expects that power is distributed unequally;
\item 
\textbf{Individualism versus collectivism (IDV):} The degree to which people within a society collaborate with each other;
Thus, in highly individualistic societies would encourage individual authority, achievement, and give the power to make individual-decision.
\item 
\textbf{Masculinity versus Femininity (MAS):} The degree to how the social gender roles are distinct;
\item 
\textbf{Uncertainty Avoidance Index (UAI):} 
The extent of society members to feel either uncomfortable or comfortable in chaotic or confused situations;
\item 
\textbf{Long- vs. Short-term Orientation (LTO):} The degree to which people within a society are linked to its own past while dealing with the present and the future challenges; and 
\item 
\textbf{Indulgence versus Impulses (IND):} The degree to which people within a society have fun and enjoy life without restrictions and regulations.
\end{itemize}

In this theory, each country has a numerical score within the above dimensions  
to define the society of this country.
The score runs from 0-100 with 50 as an average.
The Hofstede's rule for the score is that if a score is above the average, the culture relatively high on that dimension.
For example, the Saudi Arab score of the uncertainty avoidance index dimension is (80) and it is considered high.

As this research used the Saudi Arabian context for the pilot study, let us present an example of scores for this country as per \cite{hofstede2010cultures}.  
The score of power distance is very high (95), which means that Saudi people accept authority's decisions without any justifications and expect to be told what to do.
With a score of 25, Saudi Arabia is considered a collective society where people belong to group that takes care of them.
For the masculinity dimension, a score of 60 indicates that conflicts are resolved by letting the strong win and high authorities are supposed to be decisive. 
Saudi Arabia scores 80 for uncertainty avoidance which means that Saudi people prefer avoiding uncertainty situations.
The low score (36) of long-term indicates that establishing truth is the main concern as well as respect for traditions. 
Lastly, Saudi Arabia with a score of 52 does not clearly show any preference of indulgent or restrained.

\begin{figure}[ht!]
  \centering
  {\includegraphics[width=1\columnwidth]{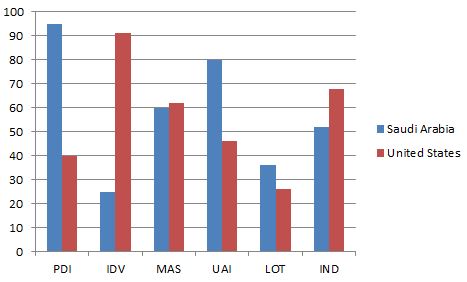}} 
  \caption{Hofstede's Culture Scores for Saudi Arabia and the United States}
  \label{fig:culture_score}
 \end{figure}

As our study argues that most of the RE standards are driven by the United States culture and the adaptation of the RE process requires considering the cultural differences, let us present a brief comparison between Saudi Arabia and the United States as per Hofstede's score (see Figure \ref{fig:culture_score}).
In short, Saudi Arabia is remarkably different from the United States within three dimensions: PDI, IDV, and UAI.
For the PDI, Saudi Arabia score is much higher than the United States (40). 
This may indicate that high authority in Saudi Arabia may have more privileges over the subordinates.
In contrast, the United State's scores very high (90) on the IND dimension, which indicates that the United States society is more individualist than Saudi Arabia.
Finally, the score of UAI in Saudi Arabia is high, which may express the difficulties of coping with unexpected situations. 

\section{\uppercase{Conceptual Model and hypotheses}}
\label{sec:Conceptual_Model}

Table \ref{tab:activities1} presents our conceptual model.
Each {\textquotedblleft}?{\textquotedblright} in the table means that the corresponding cultural dimension might have a major influence on the corresponding RE activity, whereas each {\textquotedblleft}X{\textquotedblright} means that the corresponding cultural dimensions might have less/no influence on the corresponding RE activity.
Referring to the Hofstede's scores, we believe that three of Hofstede's  cultural dimensions (PDI, IDV, and UAI) have major impacts on the practice of RE process in Saudi Arabia, while the other dimensions might have a minor impact.
Therefore, we formulated the following hypotheses (H): 
\begin{itemize}
    \item
\textbf{H1:} The high value/score of  PDI in Saudi Arabia significantly impacts on the RE activities.
 \item
\textbf{H2:} The high value/score of IDV in Saudi Arabia significantly impacts on the RE activities.
 \item
\textbf{H3:} The high value/score of UAI in Saudi Arabia significantly impacts on the RE activities.
 \item
\textbf{H4:} The slightly low value/score of MAS in Saudi Arabia may impact on the RE activities.
 \item
\textbf{H5:} The slightly high value/score of LTO in Saudi Arabia may impacts on the RE activities.
 \item
\textbf{H6:} The IND dimension has no impact on the RE activities in Saudi Arabia. 
  \end{itemize}

\begin{table}[ht!]
\centering
\caption{Influence of Cultural Dimensions on RE Activities}
\label{tab:activities1}
 \resizebox{1\columnwidth}{!}{%
\begin{tabular}{p{2cm}ccccccc}
		\hline 
		\multirow{2}{*}{\textbf{RE Activity}}& \multicolumn{6}{c}{\textbf{Cultural Dimensions}}\\
		
		 			  & PDI & IDV & MAS & UAI & LTO  & IND  \\ 
		\hline 
		Elicitation & ? & ? & ? &?  &  ? & X \\ 
		%\hline 
		Analysis& ? & ? & ? & ? &  ?& X\\ 
		%\hline 
		Specification & X & X & X &?  &? & X\\ 
		%\hline 
		Validation & ? & ?  & ?  & X & ? &X \\ 
		%\hline 
		Management & ? & ? &X  & X &X  & X \\ 
		\hline  
	\end{tabular}
 }
\end{table}

\section{\uppercase{Experts' View: Case Study}}
\label{sec:study}

\noindent 
We conducted a qualitative study through a pilot case study with interviewing Saudi software practitioners and academics.
Qualitative research is used to investigate a social behavior from the view of the study participants with attempting to achieve a holistic understanding of the situation. 
The study was conducted following the process suggested by \cite{runeson2012case}, where
the process contains the following steps: 
First, we defined the research questions and contracted the study design.
Then, we developed the case study procedure and protocol for data collection. 
Next, the first author conducted in-depth interviews in Saudi Arabia. 
Finally, the data analysis procedure was applied to the data.

The pilot case studies provide a powerful mechanism to study the aspects that support or prevent innovative behavior in software engineering practice  \citep{monteiro2016innovative}. 
Here, we used a pilot case study to understanding the Saudi cultural influence on RE process.% 
The following research questions were formulated:
\begin{itemize}[label=] 
\item 
\textbf{RQ1}: What requirements engineering activities are adopted by the Saudi requirements engineers?
\item 
\textbf{RQ2}: What cultural aspects and traditions influence the requirements engineering practice in Saudi Arabia?
\end{itemize}

\subsection{Data collection}

To collect the data, we interviewed eight participants: six software practitioners, one senior academic, and one lecturer in a Saudi university.
The snowballing technique was used to recruit the participants.
Participants were asked to propose other persons who could probably contribute to the study.
\cite{neuman2002social} suggests that snowballing sampling is often used if the sample size is small or if there is a preference to select well-informed participants.
In addition, the snowball recruiting technique was beneficial for this study because social networking, in a collectivist society, is the main source of information \cite{hofstede2010cultures}.

The practitioners interviewees were selected based the following criteria: 
\begin{itemize}
\item 
The participants had to be involved in the RE process during systems development; 
\item
Working in a Saudi software company; % 
\item 
Be employed in a medium- to a large-sized organization; and
\item
Engaged in eliciting the requirements from Stakeholders.
\end{itemize}

The academics interviewees were selected based the following criteria: 
\begin{itemize}
\item 
The participants had to be involved in the RE process during systems development; and
\item
Working in a Saudi university.
\end{itemize}

The social nature of this study leaded us to adopt semi-structured interviews.
The questions were mixed of open-ended and close-ended types including 30 questions, focusing  on 
the RE process adopted in Saudi Arabia and the cultural factors influence the RE process from Saudi Arabian context.
When the participants accepted the invitation to participate, the following documents were sent to them one day before the face-to-face interview:  the interview questions, the explanatory statement and the consent form. 
The average duration of the interviews was about 90 minutes.
Some interviews held in Arabic where other in English, based on the interviewees' preferences. 

 \subsection{Data Analysis}

A thematic analysis method was used to analyze data from the interview transcripts, following 
the guidelines described by \cite{braun2006using}.
As per \cite{marks2004research}, the benefit of thematic analysis is that it allows to combine analysis of the frequency of codes with analysis of their meaning in context, to overcome the major criticisms of content analysis. 
 The process of thematic data analysis goes through six phases:
 \begin{enumerate}% 
 \item Familiarization with data;
 \item Generating initial codes;
 \item Searching for themes;
 \item Reviewing themes;
 \item Defining and naming themes; and
 \item Producing a report.
 \end{enumerate}
All interviews were transcribed, and the interviews conducted in Arabic were also translated to English. All recordings were reviewed and coded manually using coding and pattern matching techniques \cite{miles1984qualitative}.  
Codes that emerged in each interview were compared. 
Next, we analyzed the interview data to identify common points and ideas.
Finally, we grouped the emerged 16 themes into two main groups: RE aspects and cultural aspects.

\subsection{Validity and Reliability}

Validity and reliability of research aim to ensure the trustworthiness of a study.
We considered the three validity aspects of exploratory research recommended by \cite{yin2013case}: 
First, we constructed this study through in-depth reviewing of the literature. Second, we reviewed some of the participants' software documentation that obtained during the interview as a second source of evidence. 
Finally, we established a chain of evidence from the beginning of the search through selecting the study interviewees, using interview process and transcribing and interpreting the data.
External validity achieved by using Braun's guidelines for coding and analyzing the data \cite{huberman1985assessing}. 
The reliability achieved by documenting a rich description of the research method and the interview protocol \citep{neuman2002social}.

%=======================================================================
\section{\uppercase{Results and Discussion}}
\label{sec:Result}

\subsection{Requirement Engineering Process}
\label{sec:RE_practice}

To collect the responses to the first research question, the interviewees were  asked to describe the RE processes performed in all software development projects they conducted. 
 From the participants' responses, we identified core aspects that describe the state of RE practice in the Saudi context, related to the 5 RE activities (see Section~\ref{sec:Bckground_RE}).

\subsubsection{Requirements Elicitation}

Requirements elicitation was named by all participants as an important part of the process.
The interviewees generally elicit requirements  (1) either from software stakeholders directly, or (2) by searching about the problem domains, or (3) by asking stakeholders for examples of similar systems.
Four requirement elicitation techniques were noted: interviews, focus groups, observations and prototypes.

% The most common source of the requirements was software stakeholders, such as top managers, end users, software owner or software supplier.
% The interviewees reported that the they fist meet up with customer's top managers or the decision makers to understand their views as well as to identify the main business defects.
The interviewees believed that it was important to understand the vision of the top managers or software owner before eliciting the requirements from the end users, especially for a new system.
The goals, scope, objectives and main issues from high authority people were used as guidelines to develop software that meets the company's vision, where the end users' needs are less important. 

Almost 80\% of the  interviewees agreed that a face-to-face interview was the best method used for eliciting customers' requirements.
Adopting interview as an elicitation method assisted the study participants to
\begin{itemize}
\item %$(1)$ 
acquire a comprehensive knowledge about users' requirements;
\item %$(2)$ 
verify the requirements during the session;
\item  %$(3)$ 
help clients to express their needs freely;
\item  %$(4)$  
minimize the time to establish a trust;
and 
\item %$(5)$~
build a good relationship with the customers.
\end{itemize}

One participant reported one serious problem in implementing software sponsored by the government:
requirements were identified by representative stakeholders from the government and no end users participated the elicitation session.

\subsubsection{Requirement Analysis}

Another RE activity reported by almost all participants was requirements analysis. 
The main objective of this activity is to $(1)$ detect and resolve conflicts and $(2)$ to know how the requirements should be implemented.

 The participants expressed that analyzing the requirements not only involve examining the software functionality, but also the technical, business and cultural aspects. As an example of the business aspect was that an end user may request to delete records in the business sheet, whereas according the international financial and accounting standard, such records cannot be removed. An example of technical aspects was that an end user might request a feature that threatens the company's security. For cultural aspects, an end user might ask for changing the interface language.     
To analyze the requirements, two interviewees used a scenario-based analysis method, which is a description of the steps needed to complete a given task. 

\subsubsection{Requirement Validation}

As indicated by interviewees, requirements validation was not an independent phase, as validation activities are either embedded in the requirement analysis phase, or not conducted at all.

\subsubsection{Requirement Documentation}

All participants from industry did not adopt any recommended requirements documentation standard and did not follow any guidelines in constructing the documentation template.
Instead, the requirements documentation was structured based on the requirement engineer's experience or following a documentation template used for some previous projects.
The participants gave the following three reasons for not adopting any requirements documentation standards:
\begin{itemize}
\item
The requirement engineer and the client are satisfied with the current requirements documentation;
\item
None of the existing documentation standards has been released in an Arabic requirements documentation version;  
translating  or using English documentation standards was not convenient.   
\item
The software teams are highly qualified (by their own definition) and, therefore, there is no need for a high level description of the process.
\end{itemize}
However, from the point of view of academic participants, 
following the international documentation standards (e.g., IEEE) with a customization to the company standard is seen as a more preferable solution than structuring the software documentation based on the requirement engineer's experience.

\subsubsection{Requirement Change Management}

Interestingly, the participants did not perceived requirements change management as a part of the RE process.
2 out of 5 participants from the industry adopted tools or techniques for tracking requirements changes.
This demonstrates that adaptation of the good practices and standards of RE within software development  is still in progress in Saudi Arabia.
The findings concur with \cite{alnafjan2012empirical}.

\subsection{Cultural Influences}
\label{sec:SaudiCulture}
While analyzing the collected data, we identified 10 cultural aspects influencing the RE practices:
\begin{enumerate}
\item Deference to elderly people and to high authorities;
\item Autocratic decision-making;
\item Limited trust;
\item Belief in expertise;
\item Relationships;
\item Empathy with client;
\item Letting the strongest win;
\item Gender segregation;
\item Dress code; and
\item Using English for requirements documentation.
\end{enumerate}
We grouped these points into 5 cultural categories: We cover 4 out of 6 Hofstede's dimensions (power distance, uncertainty avoidance, collectivism, masculinity) and introducing a new category to present specific cultural factors that cannot fit the Hofstede's dimensions. 
Based on the results of the conducted interviews, 
Table~\ref{tab:activities} illustrates the influence of Saudi Arabia national culture on the core activities within the RE process. % 
Each \checkmark in the Table~\ref{tab:activities} means that the corresponding RE activity is influenced by the corresponding cultural dimension.
In this model, we excluded two of the Hofstede's cultural dimensions and added a specific cultural factors.
we excluded the LTO dimension because of our data did not explore any influences or dependencies of this dimension on the Saudi RE process and we excluded the indulgence dimension because it does not have any connection with the RE process.
Also, we added the specific cultural factors as it emerged during analyzing the study participants' interviews.
We named it specific because it is exclusively distinct to the Saudi culture. 

\begin{table}[ht!]
\centering
\caption{Influence of Saudi culture on RE Activities}
\label{tab:activities}
\resizebox{1\columnwidth}{!}{%
\begin{tabular}{p{2cm}ccccccc}
		\hline 
		\multirow{2}{*}{\textbf{RE Activity}}& \multicolumn{5}{c}{\textbf{Cultural Dimensions}}\\
		
		 			  & PDI & IDV & MAS & UAI  & Specific \\ 
		\hline 
		Elicitation & {\checkmark} & {\checkmark}  &   &{\checkmark}  &  {\checkmark} \\ 
		%\hline 
		Analysis& {\checkmark} &   & {\checkmark}  & {\checkmark} &   \\ 
		%\hline 
		Specification &  & {\checkmark} &  &   & {\checkmark} \\ 
		%\hline 
		Validation & {\checkmark} &  & {\checkmark}  &   \\ 
		%\hline 
		Management & {\checkmark}  & {\checkmark} &    & \\ 
		\hline  
	\end{tabular}
 }
\end{table}

\subsubsection{Power Distance Index}

As mentioned in Section~\ref{sec:Background_Hofstede}, the score of power distance index (PDI) in Saudi Arabia is defined as 95 (very high).
The following three points fit this dimension: deference to elderly people and to high authorities,
autocratic decision-making, and lack of trust.

\textbf{Deference to elderly people and to high authorities:} 
As per \citep{hofstede2010cultures},  in the countries with a high PDI, showing respect for parents and older people is a basic and lifelong virtue. 
The data collection highlighted the fact that due to the Saudi Arabian culture, demonstrating respect was habitually practiced.
A total of five participants reported that deference of customers, especially elderly and high authorities, is seen by them as a standard practice.
One of the participants stated that \textit{{\textquoteleft}most of the time, I cannot argue with elderly or high authority employees{\textquoteright}}.

\textbf{Autocratic decision-making:} \cite{hofstede2010cultures} stated that \textit{{\textquotedblleft}power distance will affect the degree of centralization of the control and decision-making structure and the importance of the status of the negotiators{\textquoteright}}.
 In that context, the analysis of the collected data demonstrated that the decision-making was significantly consultative and autocratic rather than democratic.
 The reason for this is a centralized decision-making in Saudi Arabia, i.e., all decisions are taken by high authorities.

 Three participants believed that \textit{{\textquoteleft}The manager will agree on what he thinks it is beneficial to the business. All about business. If it is beneficial, he will accept it{\textquoteright}.} and therefore the study participants accept and agree one what is decided by the top managers.
However, one participant claimed a critical drawback to the centralized decision as \textit{{\textquoteleft}If he [decision maker] is satisfied, we do not care about the end users' needs{\textquoteright}}.

\textbf{Limited trust:} According to \cite{hofstede2010cultures}, high PDI cultures suffer from lack of trust. A total of five participants mentioned that lack of trust in both Saudi and non-Saudi requirements engineer was a problematic issue.
During the elicitation session, clients tended to not openly discuss the main core of the business issues, instead, 
clients may ask the requirements engineer about: the number of projects he was involved in, years of experience, race, the education and technique background, etc.  % 
Surprisingly, the level of trust is normally established based on the requirements engineering's nationality or based on the strength of the relationship that built during the RE phase. 
One participant commented that some nationalities were highly trusted just because Saudi people believe in them. 
Also, three participants agreed that establishing a good relationship with customers significantly increased the trust level. 
 However, if the level of trust is not maintained on the early stage of the software development life cycle, the RE process might end up with an absolute failure to achieve its purposes. 

\subsubsection{Uncertainty Avoidance}
 
The score of uncertainty avoidance in Saudi Arabia is defined as 80 (very high), cf. Section~\ref{sec:Background_Hofstede}. 
According to Hofstede, in high uncertainty cultures, strong belief in expertise and their knowledge is common \cite{hofstede2010cultures}.

The professional informers assume that  Saudi customers accept their recommendations and suggestions not because they are experts in the field, but because these recommendations are important and valuable. 
According to the interviewees, presenting facts about the best way to solve the problem can change software stockholders' mind easily. 
However, this expresses only the view point and self-estimates of requirements engineers rather than provides a complete real picture, where the belief in expertise (including the belief in one's own expertise) plays a large role.

\subsubsection{Collectivism}

Saudi Arabia is considered a collective society (score 25, cf. Section~\ref{sec:Background_Hofstede}), where two following points play an important role: building a relationship and empathy with users.

 \textbf{Building a relationship:} 
 In a collectivism cultures, building a relationship in workplaces is very important, similar to family relationships.
Interviews, as a face-to-face method to elicit the requirements, permit the software engineers to create an emotional relationship with the client, which was indicated as a positive culture aspect.
Six participants highlighted that building a good relationship means having an in-depth understanding of users' problem domain as well as establishing a certain level of trust.
For example, during the requirements elicitation building a relationship with the clients can be used as facilitator to overcome barriers (e.g, lack of user involvement) or challenges in general such as project over-budget.

\textbf{Empathy with users}: Achieving user satisfaction was continually mentioned as the reason to accept new changes, which can be explained in terms of feeling empathy with the client to fulfill his needs.
Four participants agreed that \textit{{\textquoteleft}He is a client. We cannot make him unhappy.
The client should be satisfied{\textquoteright}}.

\subsubsection{Masculinity}
Solving conflicts by \textbf{letting the strongest win} was often mentioned in the transcript.
As per Hofstede' theory, the resolution of conflicts in a masculine society (Saudi Arabia is considered a masculine society with score 60) is solved by letting the strongest win \cite{hofstede2010cultures}.

In Saudi context, autocratic behavior was indicated when the interviewee spoke about solving conflicts between end users and high authorities.
The findings showed that most of the conflicts were resolved by agreeing with the top managers' preferences rather than giving end users an opportunity to justify this need.
Moreover, we find that this cultural aspect is highly related to the PDI dimension.
In high PDI dimension, such as Saudi Arabia, the decision is made based on the higher authorities' opinion.

\subsubsection{Specific Cultural Factors}

The study identified three points that are very for Saudi culture: 
gender segregation, dress code, and English language, where the first two can also be seen as a very special application of masculinity dimension. 

\textbf{Gender segregation:}   
A total of five participants asserted that communicating with Saudi women during the RE process, even during the elicitation activity, was not an issue.
The study participants approached the female end users in their workplace.
Even though the workplace is separated into male and female sections, the professionals' participants elicited the requirements through either face-to-face meetings or using communication applications such skype or e-mail.  
However, only one participant reported that gathering the requirements from female was difficult due to shyness and self-respect of the Saudi women.

\textbf{Dress code}: 
Some participants highlighted that wearing a suit for meeting with clients is very important. They perceived this rule as a cultural aspect.

\textbf{English language}: Even though the official written language in Saudi Arabia is Arabic, all participants indicated that software documentation is usually written in English, especially for government businesses.
 Since Saudi people have diverse levels of the speaking and writing English language skills,  adopting English language to communicate and document the requirements might introduce miscommunication and misunderstanding between the clients and requirements engineers  and, therefore, it might affect the overall RE process.

\begin{figure*}[ht!] 
   \centering
  {\includegraphics[width=1.6\columnwidth]
{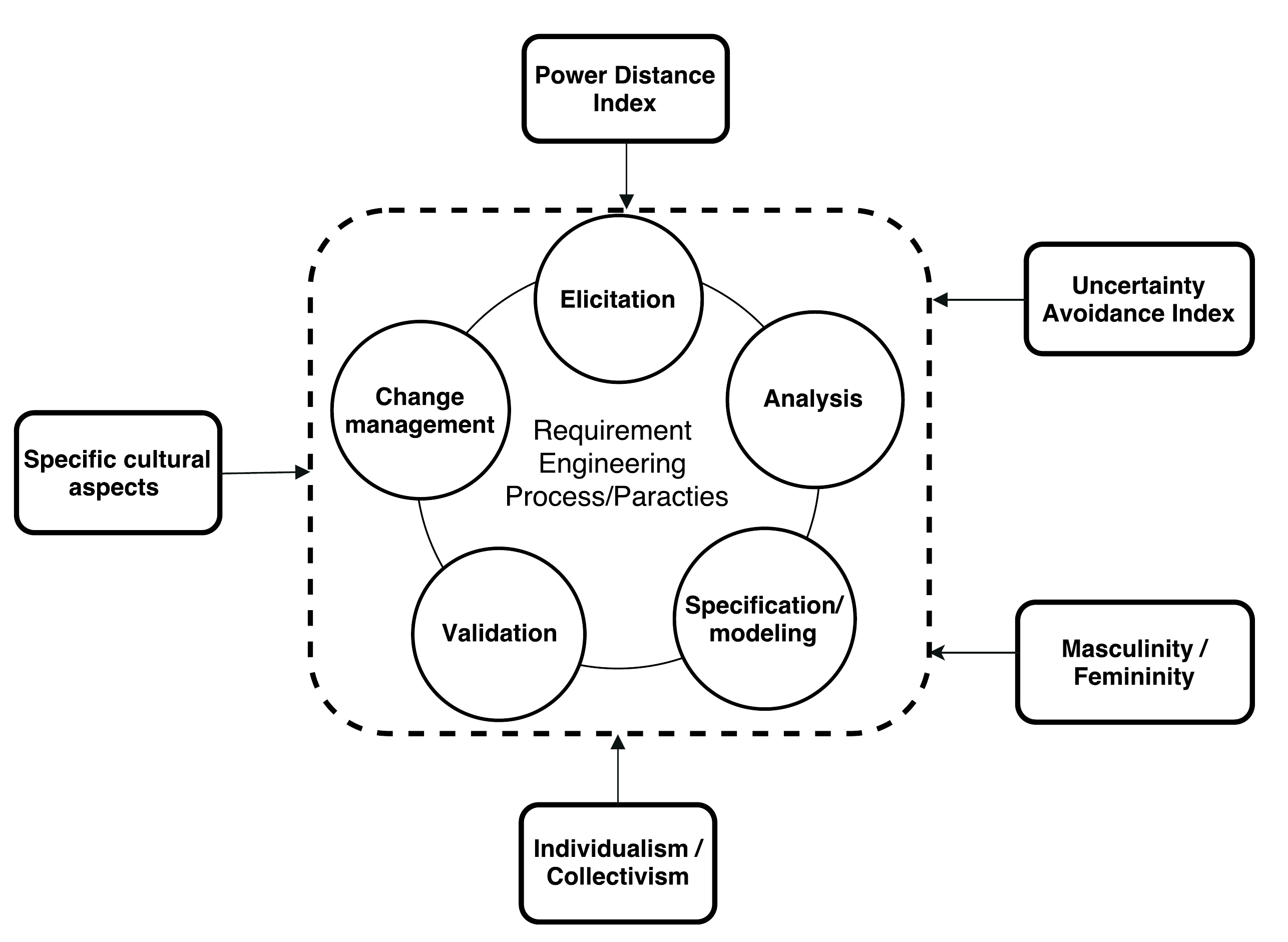}} 
%  
%
% ^.
  \caption{Framework of RE process in a cultural context}
  \label{fig:model1}
 \end{figure*}
 
%=======================================================================
\section{\uppercase{Proposed Framework}}
\label{sec:model}
\noindent 
Based on the expert participants' overview,
we designed a framework of the national cultural influence on the practice of RE process as presented in Figure~\ref{fig:model1}. 
We are planning to replicate the study and compare the results with several cultures, in order to propose a generalized framework that identifies the influence of national culture on the practices of RE process.
For this purposes, we extended and refined the initial model (presented in Table \ref{tab:activities1}) to explore the influence of the cultural background on the RE process within other country/culture context, as presented in Table \ref{tab:activities2}. 
Thus, based on the Hofstede's country/culture score for a particular dimension, we can explore (1) what RE activities have to be adapted and (2) which of the culture factors might have an influence on the RE process.

\begin{table}[ht!]
\centering
\caption{Influence of Cultural Dimensions on RE Activities}
\label{tab:activities2}
 \resizebox{1\columnwidth}{!}{%
\begin{tabular}{p{2cm}ccccccc}
		\hline 
		\multirow{2}{*}{\textbf{RE Activity}}& \multicolumn{7}{c}{\textbf{Cultural Dimensions}}\\
		
		 			  & PDI & IDV & MAS & UAI & LTO  & IND & Specific \\ 
		\hline 
		Elicitation & ? & ? & ? &?  &  X & X& ? \\ 
		%\hline 
		Analysis& ? & ? & ? & ? &  ?& X& ?\\ 
		%\hline 
		Specification & X & ? & X &?  &? & X& ?\\ 
		%\hline 
		Validation & ? & ?  & ?  & X & ? &X & ?\\ 
		%\hline 
		Management & ? & ? &X  & X &X  & X&? \\ 
		\hline  
	\end{tabular}
 }
\end{table}

Based on the initial results, we formulated a number of new  hypotheses for our future work: 
\begin{itemize}
    \item
\textbf{H1:} The high PDI has an impact on the practice of RE activities.
 \item
\textbf{H2:} The low IDV has an impact on the practice of RE activities.
 \item
\textbf{H3:} The high UAI has an impact on the practice of RE activities.
 \item
\textbf{H4:} The high/low MAS has an impact on the practice of RE activities.
 \item
\textbf{H5:} The high LOT cultural dimension has an impact on the practice of RE activities.
 \item
\textbf{H6:} The IND dimension has no impact in the practice of RE activities. 
\item
\textbf{H7:} There are specific cultural values impact the practice of RE activities, within each country/culture context. 
\end{itemize}
% 

%=======================================================================
\section{\uppercase{Related Work}}
\label{sec:RELATED}

\noindent To analyze cultural aspects, many social psychologists propose theoretical frameworks that illustrate the key cultural factors that distinguish one culture from another, cf. e.g., \cite{hofstede2010cultures}, \cite{trompenaars2011riding}, and \cite{hall1989beyond}. 
However, from our best knowledge, our work is the first one focusing on the cultural aspects within the RE process in Arabian culture. 
In our earlier works, we analyzed individual and social requirement aspect of software development, cf. \cite{alharthi2016individual,alharthi2015requirements}, as well as requirements for global product development, focusing on the diversity, which comes from the differences in the regulations, laws and cultural aspects  for different countries or organizations, cf. \cite{spichkova2015structuring,spichkova2015requirements}. In our current work, we went further and investigated cultural aspects in details, focusing on the Arab culture.

Several studies have investigated the impact of culture and cultural differences on RE processes. For example, \cite{damian2003re} analyzed the challenges faced by stakeholders during the RE process in distributed multi-site organizations. By conducting in-depth interviews, the study showed that culture and cultural differences have negative impacts on RE activities. However, the existing studies do not propose any solution that can be generalized and applied in the future. The goal of our work is to solve this problem. 

\citet{thanasankit2002requirements} conducted a study on whether the specifics of Thai culture might have influences on the RE process, in particulate, the concept of power distance.
According to the authors, the communication and decision-making process in Thailand is
influenced by the unequal power distribution and the level of uncertainty. 
Based on Thai cultural aspects, the researchers suggested some strategies to improve RE practices.
These strategies focus on Thai culture only and cannot be generalized. However, the results of the study concur with our hypothesis that the cultural influence on the RE process might be significant and it should be taken into account to improve the RE practices.
In contrast to \citet{thanasankit2002requirements}, our research adopted the six dimensions of the Hofstede's cultural theory on the RE process and the results of the pilot case study demonstrated the  feasibility of our approach.

 \citet{isabirye2008model} evaluated the existing requirements elicitation techniques and methodologies in the context of rural culture in South Africa, coming to the conclusion that the stakeholders' culture plays a significant role in undertaking the RE process, which concurs with our hypotheses. 
 \cite{Borchers} analyzed the impact of culture on the software engineering techniques  in Japan, India, and the United States.  The study also demonstrated that cultural differences have a great influence on software engineering process.  
 In our work, we go further and focusing on the development of the corresponding framework.
%   

%=======================================================================   

\section{\uppercase{Conclusion}}
\label{sec:conclusion}
\noindent RE process is highly sensitive to cultural aspect and requires extensive communication skills.
Several studies have recognized that a success RE process does not only depend on the technique or method used but also depends on understanding the individuals' behavior and culture.
In this paper, we investigate the influence of cultural background on performing the RE process from a conservative culture, Saudi Arabia.    

We conducted interviews with software practitioners and software engineering researchers in Saudi Arabia.
Using thematic analysing technique with eight participants, this study identifies 10 cultural factors play a role on the RE practice and 5 RE aspects from the Saudi context. 
As a result, we introduce a model of RE processes with a cultural context. 
In this model, we map the Hofstede's cultural theory to the core parts of the RE process to investigate the possible dependencies and influences. 

Hofstede's culture dimension allows us to find relevant correlation regarding the impact of the cultural background on RE practice. 
The result of this study, therefore, would help the Saudi RE practitioners and other RE practitioners, having the similar national cultural profile, to identify in advance the potential cultural implications that might be faced.

\textbf{Future work:} 
As the next steps, we are going to expand the scope of the study  
(1) to cover other countries, (2) to gather the information on the software stakeholders' view points, as they are a very important actors within the RE process, (3) to cover not only small-to-medium companies but also a large ones, as this might provide additional aspects. 
We are planning to extend this work to a large number of participants and comparing the impact within several cultures.

\section*{{Acknowledgements}}
We would like to thank participants for kindly accepting to participate in this research project. We appreciate their cooperation and engagement.\\

\noindent 
The first author is supported by a scholarship from Taibah University in Saudi Arabia.
% 

% \vfill
\bibliographystyle{apalike}
{\small

%\bibliography{Master_conf.bib}
}

\end{document}